\documentclass[showpacs,twocolumn,amsmath,amssymb]{revtex4}
\usepackage{graphicx}
\usepackage{psfrag} 
\usepackage{epsfig}    
\usepackage{rotating}  

\begin{document}

\title{Dynamics of Non-Conservative Voters}
\author{R. Lambiotte$^{1}$\email{Renaud.Lambiotte@uclouvain.be}}
\author{S. Redner$^{2}$}
\affiliation{
  $^1$ INMA, Universit\'e catholique de Louvain,  4 avenue Georges Lemaitre,
B-1348 Louvain-la-Neuve, Belgium \\
$^{2}$Center for Polymer Studies and Physics  Department, Boston University, Boston,
MA~ 02215 USA}


\begin{abstract}
  We study a family of opinion formation models in one dimension where the
  propensity for a voter to align with its local environment depends
  non-linearly on the fraction of disagreeing neighbors.  Depending on this
  non-linearity in the voting rule, the population may exhibit a bias toward
  zero magnetization or toward consensus, and the average magnetization is
  generally not conserved.  We use a decoupling approximation to truncate the
  equation hierarchy for multi-point spin correlations and thereby derive the
  probability to reach a final state of $\uparrow$ consensus as a function of
  the initial magnetization.  The case when voters are influenced by more
  distant voters is also considered by investigating the Sznajd model.
\end{abstract}

\maketitle


It often happens that individuals change their attitudes, behaviors and/or
morals, to conform to those of their acquaintances.  Perhaps the simplest
description of this conformity is the voter model \cite{L99}, where each node
of a graph ({\it i.e.}, the social network) is occupied by a voter that has
one of two opinions, $\uparrow$ or $\downarrow$.  In the voter model, the
population evolves by: (i) picking a random voter; (ii) the selected voter
adopts the state of a randomly-chosen neighbor; (iii) repeating these steps
{\it ad infinitum} or until a finite system necessarily reaches consensus.
Figuratively, voters have no self confidence and merely follow one of their
neighbors.  With this dynamics, a voter changes opinion with a probability
$p_f$ that equals the fraction $f$ of disagreeing neighbors.  This
proportionality rule leads to the conservation of the average opinion in the
system, a feature that renders the voter model soluble in all dimensions
\cite{L99,K02}.  However, the specific rule $p_f=f$ is one among many
possible and socially plausible relations between $p_f$ and $f$.

\begin{figure}[ht]
\includegraphics[width=0.46\textwidth]{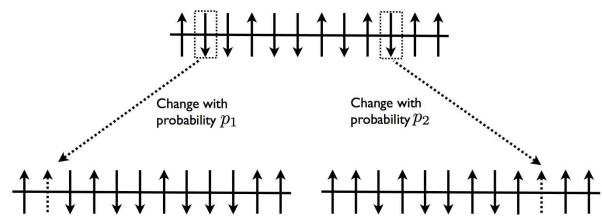}
\caption{Update illustration.  A random voter changes state with probability
  $p_1$ if it has 1 disagreeing neighbor (left), and with probability $p_2$
  if it has 2 disagreeing neighbors (right).}
\label{model}      
\end{figure}

In this work, we generalize the voter model so that $p_f$ depends
non-linearly on $f$.  Non-linear voter models have been discussed previously,
primarily by numerical simulations in two dimensions \cite{nonlinear}.  Here
we focus on one dimension, where the range of possibilities for the
non-linearity is limited.  In one dimension, a voter may be confronted by 0,
1, or 2 disagreeing neighbors.  It is natural to impose $p_0=0$, so that no
evolution occurs when there is local consensus.  Then the most general
description of the system requires two parameters, $p_1$ and $p_2$
(Fig.~\ref{model}).  One parameter, which we choose to be $p_1$, determines
the overall time scale and is thus immaterial.  The only relevant parameter
then is $\gamma = p_2/p_1$.  When $\gamma=2$, one recovers the classical
voter model.  When $\gamma>2$, the combined effect of two neighbors is more
than twice that of one neighbor.  Equivalently, voters can be viewed as
having a conviction for their opinion and that strong peer pressure is needed
to change opinion.  As $\gamma \rightarrow \infty$, voters only change
opinion when are confronted by a unanimity of opposite-opinion voters
\cite{unanimity}.  In contrast, when $\gamma<2$, one disagreeing neighbor is
more effective in triggering an opinion change than in the classical voter
model.  When $\gamma=1$, one recovers the {\em vacillating\/} voter model
\cite{rapid,morris} where voters change opinion at a fixed rate if either 1 or 2 of
their neighbors disagree with them.  Finally, $\gamma<1$ corresponds to a
``contrarian'' regime where a voter is less likely to change opinion as the
fraction of neighbors in disagreement increases.

We term $\gamma$ the conviction parameter and view voters as having or
lacking conviction if $\gamma>2$ or $\gamma<2$, respectively.  When
$\gamma>2$, the population is quickly driven to consensus, while when
$\gamma<2$, there is a bias toward the zero-magnetization state with equal
densities of voters of each type.  In the former case, the collective
evolution is more extreme than the evolution of an individual.
In the latter case, collective evolution is hindered and the approach to final
consensus is anomalously slow.  In both cases, we can determine the {\em exit
  probability}, namely, the probability the system ultimately reaches
$\uparrow$ consensus as a function of the initial composition of the
population.

In the framework of the Ising-Glauber model \cite{IG}, the transition rate of a
voter at site $i$, whose states we now represent by $\sigma_i=\pm 1$, is
\begin{eqnarray}
\label{w}
w(\{\sigma\}\!\!\rightarrow\!\!\{\sigma^{'}\}_i)&\!\!=\!\!&
\gamma+2-\gamma\sigma_{i}(\sigma_{i-1}+\gamma\sigma_{i+1})~~~~~~~~\cr
&&~~~~~~~~+  (\gamma-2 )\sigma_{i-1}\sigma_{i+1}, 
\end{eqnarray}
with $\{\sigma\}$ denoting the state of all voters, $\{\sigma'\}_i$ the state
where the $i^{\rm th}$ voter has flipped, and where an irrelevant overall
constant has been absorbed in the time scale.  The fact that the $\sigma_{i}
\sigma_{i+1}$ and the $\sigma_{i} \sigma_{i-1}$ terms have the same
coefficient is due to the left/right symmetry of the dynamics.  Note also
that for $\gamma=2$, the $\sigma_{i-1}\sigma_{i+1}$ term vanishes, and the
equation of motion of the classic voter model is recovered.  As shown in
Glauber's original work \cite{IG} and as we show below, the
$\sigma_{i-1}\sigma_{i+1}$ term couples the rate equation for the mean spin
to 3-body terms.

Because the average magnetization is not conserved, the model is not
exactly soluble.  Nevertheless, we can construct an approximate solution by
truncating the hierarchy of rate equations of higher-order correlation
functions by a simple decoupling scheme.  The mean spin,
$s_j\equiv\langle\sigma_j\rangle=\sum_{\{\sigma\}} \sigma_j P(\{\sigma\};t)$
evolves as
\begin{eqnarray}
\label{st}
\frac{\partial s_j}{\partial t}&\!\!=\!\!& \sum_{\{\sigma\}}\sigma_j\Big[\sum_i w(\{\sigma'\}_i
\!\!\rightarrow\!\! \{\sigma\})\, P(\{\sigma'\}_i;t) \nonumber \\
&~&~~~~~~~~~~~~-w(\{\sigma\}\!\! \rightarrow\!\! \{\sigma'\}_i) \, P(\{\sigma\};t)\Big] .
\end{eqnarray}
After straightforward steps, Eq.~\eqref{st} reduces to
\begin{eqnarray}
\label{eqS}
\frac{\partial s_j}{\partial t} = 2\gamma(s_{j+1}+s_{j-1}) - 2 (\gamma + 2 )s_j \cr
  - 2 (\gamma - 2 ) \langle\sigma_{j-1}\sigma_j\sigma_{j+1}\rangle,
\end{eqnarray}
which depends only on the mean spins $s_j$, $s_{j-1}$ and $s_{j+1}$ when
$\gamma=2$, as expected, but is otherwise coupled to higher order
correlations.  

Let us first consider the mean-field limit, where the spins of neighboring
nodes are uncorrelated.  Because of spatially homogeneity, $\langle
s_j\rangle$ are all identical and we write the magnetization as $m\equiv
\langle s_j\rangle$.  Then assuming that
$\langle\sigma_{j-1}\sigma_j\sigma_{j+1}\rangle \approx m^3$, Eq.~(\ref{eqS})
simplifies to
\begin{eqnarray}
\label{mimi}
\frac{\partial m}{\partial t}  =   2  (\gamma-2) (m-m^3),
\end{eqnarray}
which shows that the magnetization is not conserved when $\gamma \neq 2$.
The stable solutions of Eq.~(\ref{mimi}) are either consensus ($m=\pm1$) when
$\gamma>2$, or stasis, with equal densities of the two types of voters
($m=0$), when $\gamma <2$.

In one dimension, however, the mean-field assumption is not justified.  To
determine the behavior of the system when $\gamma \neq 2$, we therefore use
the approach developed in \cite{rapid} (see also \cite{MR}) to truncate the
hierarchy of equations for multi-spin correlation functions.  Consider the
rate equation for the nearest-neighbor correlation function $\langle\sigma_j
\sigma_{j+1}\rangle$:
\begin{eqnarray}
\label{ct}
\frac{\partial \langle\sigma_j \sigma_{j\!+\!1}\rangle}{\partial t} &=&
- 2 (\gamma-2)\left[\langle\sigma_{j-1} \sigma_{j}\rangle   +
   \langle\sigma_{j+1}\sigma_{j+2}\rangle\right]~~~~\cr
 &&~~~+ 2 \gamma \left[ \langle \sigma_{j-1} \sigma_{j+1} \rangle 
+  \langle \sigma_{j} \sigma_{j+2} \rangle\right]   \cr
&&~~~+ 4 \gamma  -4 (\gamma + 2 ) \langle \sigma_j \sigma_{j+1} \rangle .
\end{eqnarray}
To close this equation, we need to approximate the second-neighbor
correlation function $\langle\sigma_{j} \sigma_{j+2}\rangle$.  Consider
domain walls---nearest-neighbor anti-aligned voters---whose density is
$\rho=(1-\langle \sigma_i\sigma_{i+1}\rangle)/2$.  According to the
transition rate in Eq.~\eqref{w}, an isolated domain wall diffuses freely for
any $\gamma$.  However, when two domain walls are adjacent, they annihilate
with probability $P_{\rm a}= \gamma/(2+\gamma)$ or they hop away from each other
with probability $P_{\rm h}= 2/(2+\gamma)$.  Thus when $\gamma>2$,
$P_{\rm a}>P_{\rm h}$ and adjacent domain walls have a tendency to
annihilate, while they are repelled from each other when $\gamma < 2$.  The
interaction of two domain walls is therefore equivalent to single-species
annihilation, $A+A\to 0$, but with a reaction rate that is modified compared
to freely diffusing reactants because of this interaction.  Nevertheless, the
domain wall density asymptotically decays as $t^{-1/2}$ for any
$\gamma<\infty$, and with an interaction-independent amplitude \cite{AA}.

Since domain walls become widely separated at long times, we therefore
approximate \cite{rapid} the second-neighbor correlation function as
$\langle\sigma_{j} \sigma_{j+2}\rangle \approx \langle\sigma_{j}
\sigma_{j+1}\rangle$.  We also define $m_2 \equiv \langle\sigma_{j}
\sigma_{j+1}\rangle$ for a spatially homogeneous system.  Now the rate
equation \eqref{ct} for the nearest-neighbor correlation function becomes
\begin{equation}
\label{RE-m2}
\frac{\partial m_2}{\partial t} = 4 \gamma - 4 \gamma m_2.
\end{equation} 
For the uncorrelated initial condition, $m_2(0)=m(0)^2$, the solution is
\begin{eqnarray}
m_2(t) =  1 + \left[m(0)^2-1\right]\, e^{-4 \gamma t}~,
\end{eqnarray}
where $m(0)\equiv\langle s_j(0)\rangle $ is the average magnetization at
$t=0$. 

In a similar spirit, we decouple the 3-spin correlation function $\langle
\sigma_{j-1} \sigma_j \sigma_{j+1}\rangle \approx m m_2$ \cite{rapid} and
average over all sites, to simplify the rate equation \eqref{eqS}
for the mean spin in a spatially homogeneous system to
\begin{eqnarray}
\frac{\partial m}{\partial t}  = 2(2-\gamma) me^{-4\gamma t} (m(0)^2-1).
\end{eqnarray}
Solving this equation and taking the $t\to\infty$ limit, we obtain a
non-trivial relation between the final magnetization $m(\infty)$ and the
initial magnetization $m(0)$:
 \begin{eqnarray}
\label{m-inf}
m(\infty)  &=&  m(0)\, e^{(2-\gamma) (m(0)^2-1)/2\gamma}~ .
\end{eqnarray}
It is important to realize that the average magnetization in a finite
population does not perpetually fluctuate around this asymptotic value but
rather ultimately reaches $\pm 1$ because consensus is the only absorbing
state of the stochastic dynamics.  We characterize this approach to consensus
by the exit probability $\mathcal{E}(x,N)$, defined as the probability that a
population of $N$ voters ultimately reaches $\uparrow$ consensus when there
are initially $n=x N$ $\uparrow$ voters.  Since the density of $\uparrow$
voters is $x=(1+m)/2$ and $m(\infty)= 2\mathcal{E}(x)-1$, Eq.~\eqref{m-inf}
leads to
 \begin{eqnarray}
 \label{exit1D}
\mathcal{E}(x)  =  \frac{1}{2} \left[(2x-1) e^{2x(2-\gamma)(x-1)/\gamma} +1 \right]~.
\end{eqnarray}
The exit probability $\mathcal{E}(x)$ is independent of $N$, but has a
non-trivial dependence on the initial condition.  Similar behaviors have been
previously found in other opinion evolution models where the average
magnetization is not conserved, such as the majority rule model \cite{krap}
and the Sznajd model \cite{sznRel}.

\begin{figure}[ht]
\includegraphics[angle=-90,width=0.45\textwidth]{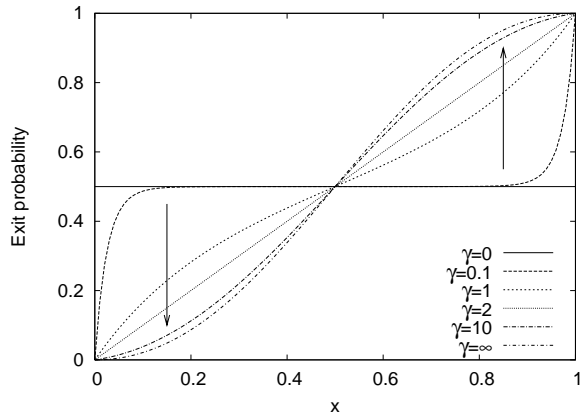}
\caption{Exit probability $\mathcal{E}(x)$ from Eq.~(\ref{exit1D}) as a
  function of the initial density of $\uparrow$ voters $x$ for different
  values of $\gamma$. The arrow indicates the position of the curves for
  increasing values of $\gamma$, namely from zero magnetization
  $\mathcal{E}(x) =1/2$ ($\gamma=0$) to more and more consensual systems.}
\label{fig2}      
\end{figure}

\begin{figure}[ht]
\includegraphics[angle=-90,width=0.45\textwidth]{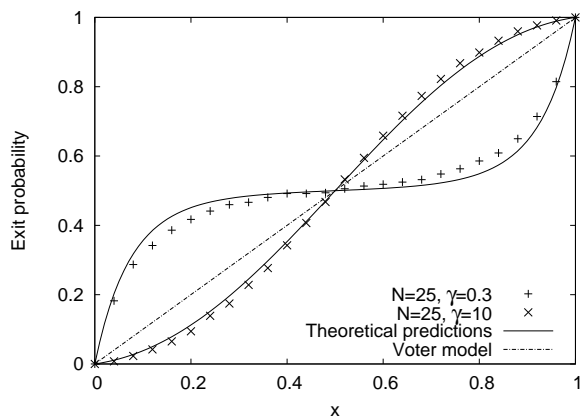}
\caption{Exit probability $\mathcal{E}(x)$ for a one dimensional system
  composed of 25 voters, for conviction parameter $\gamma=0.3$ and 10.}
\label{fig3}      
\end{figure}

The exit probability in (\ref{exit1D}) exhibits a qualitative change as
$\gamma$ passes through 2 (see Fig.~\ref{fig2}).  At $\gamma=2$, the pure
voter model result $\mathcal{E}(x)=x$, that follows from magnetization
conservation, is recovered.  When $\gamma<2$, the dynamics drives the system
toward zero magnetization before consensus is ultimately reached.  Thus the
exit probability becomes progressively more independent of the initial
magnetization as $\gamma \rightarrow 0$.  When $\gamma>2$, in contrast, the
system is driven toward consensus; in the $\gamma \rightarrow \infty$ limit,
$\mathcal{E}(x)$ reduces to
\begin{eqnarray}
 \label{asym1D}
\mathcal{E}(x) = \frac{1}{2} \left[(2x-1) e^{-2x(x-1)}+1\right].
\end{eqnarray}

We checked the validity of Eq.~\eqref{exit1D} by simulations.  To do so, we
focused on small systems ($N=25$) and directly measured the probability
$\mathcal{E}(x)$ that the population ultimately reaches a $\uparrow$
consensus when the proportion of initially $\uparrow$ voters is $x$ by
averaging over 5000 realizations of the dynamics.  The theoretical results
are in excellent agreement with our simulation results (see Fig.~3).

It is in principle possible to generalize the model to higher dimensions $d$.
However, in that case, each node is surrounded by $2d$ neighbors and $2d$
variables are therefore required, {\it i.e.}, the probabilities $p_i$ that an
individual with $i\in [1,2d]$ disagreeing neighbors changes opinion, in order
to specify completely the voters dynamics.  Such a general analysis, that
goes in the direction of the general model of contagion introduced by Dodds
and Watts \cite{watts} and would include known models for opinion formation
such as threshold models \cite{threshold}, will be considered in detail
elsewhere.  One should stress, however, that a much richer phenomenology can
occur when the number of parameters $p_i$ is increased \cite{galam}.  In the
case when each node is surrounded by four neighbors, for instance, and
within a mean-field description, it is easy to show that the equation of
evolution for the average density of $\uparrow$ voters is
\begin{eqnarray}
\label{eq2D}
\frac{\partial x}{\partial t} =   \sum_{i=1}^4 \binom{4}{i} p_i
\left[(1-x)^{5-i} x^{i} -x^{5-i} (1-x)^{i}  \right],
\end{eqnarray}
where we have again assumed that $p_0=0$.
The stable stationary solutions of Eq. (\ref{eq2D}) are either consensus (when $p_1<p_4/4$), a state of zero
magnetization (when $-3 p_1 - 3 p_2/2 +  p_3 + 3 p_4/4 <0$), or other asymmetric solutions, {\it e.g.}, when $p_1=1/2$,
$p_0=p_2=0$ and $p_3=p_4=1$, the stable solutions are $x=(5 \pm
\sqrt{5})/10$.

Let us now return to our analysis of one-dimensional systems.  The approach
developed above can also be applied to another simple opinion dynamics model
in one dimension, namely, the Sznajd model \cite{szn}.  This model is an
appealing realization of the concept of social validation, namely, that
agents are only influenced by groups ({\it e.g.}, pairs) of aligned voters
and not by single individuals.  The Sznajd is defined by the following
evolution rule: (i) pick a pair of neighboring voters $i$ and $i+1$; (ii) if
these voters have the same opinion $\sigma_i=\sigma_{i+1}$, convert the
opinion of the neighbors $i-1$ and $i+2$ on either side of the initial pair:
$\sigma_{i-1}=\sigma_{i}=\sigma_{i+1}=\sigma_{i+2}$; (iii) repeat these steps
{\it ad infinitum} or until a finite system necessarily reaches consensus.

\begin{figure}[ht]
\includegraphics[angle=-90,width=0.45\textwidth]{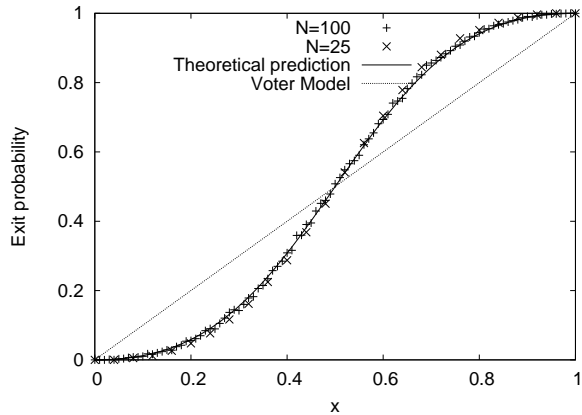}
\caption{Exit probability $\mathcal{E}(x)$ for the Sznajd model. The system
  is one dimensional and composed of 25 and 100 voters respectively. Results
  are averaged over 5000 realizations of the random process.}
\label{finalFig}      
\end{figure}

It is straightforward to show that the transition rate of a voter at site $i$
is
\begin{eqnarray}
  w(\{\sigma\}\!\! \rightarrow \!\!\{\sigma^{'}\}_i) &\!\!=\!\!&  - \frac{1}{4}
\left[ \sigma_{i}(\sigma_{i-2}+\sigma_{i-1}+\sigma_{i+1}+\sigma_{i+2})\right.~~~~~~\nonumber\\
&&~~~~~~~ \left.-\sigma_{i-1}\sigma_{i-2}-\sigma_{i+1}\sigma_{i+2}-2\right];
\end{eqnarray}
notice that the rate for changing $\sigma_i$ involves terms that include
$\sigma_{i\pm2}$.  Following the same steps as in our previous example of the
voter model with variable conviction, the resulting equations for the average
magnetization $m$ and the nearest-neighbor correlation $m_2$ for a spatially
homogeneous system are
\begin{eqnarray}
\label{eqSnaj}
\frac{\partial m}{\partial t}  =  m  -  m m_2\,,  \qquad
 \frac{\partial m_2}{\partial t} =  1 - m_4\,,
\end{eqnarray}
where we have again 
assumed that $\langle\sigma_{j} \sigma_{j+2}\rangle \approx \langle\sigma_{j}
\sigma_{j+1}\rangle$ and $\langle \sigma_{j-1} \sigma_j \sigma_{j+1}\rangle
\approx m m_2$.
Here $m_4\equiv \langle\sigma_{j-1}\sigma_j\sigma_{j+1}\sigma_{j+2}\rangle$
is the correlation between the states of four contiguous voters.  This new
term is due to the distant interactions between voters, and is also present
in the majority rule model.  Following \cite{MR}, we apply the Kirkwood
approximation to factorize the 4-point function as the product of 2-point
functions, $m_4 \approx m_2^2$; this approach has proved quite useful in a
variety of applications to reaction kinetics \cite{kirk}.  
This truncation then allows us to solve the second of Eqs.~\eqref{eqSnaj} to give
$m_2 =  \frac{e^{2t} + C}{e^{2t} - C}$,
where $C=(m(0)^2-1)/(m(0)^2+1)$, and finally to solve the equation for the
average magnetization
\begin{eqnarray}
\label{m1}
m =  \frac{e^{2t} }{e^{2t} - C} ~ \frac{2 m(0)}{1+m(0)^2}~.
\end{eqnarray}
After some straightforward algebra, we then find the exit probability
$\mathcal{E}(x)$
 \begin{eqnarray}
 \label{final}
\mathcal{E}(x)  =  \frac{x^2}{1- 2x + 2x^2}~.
\end{eqnarray}
One should stress that such a non-trivial exit probability has been observed
previously in simulations of the Sznajd model \cite{sznRel}, which clearly
shows that the Sznajd model has a tendency toward consensus and that the
prediction (\ref{final}) is in remarkably good agreement with our
simulations (see Fig.~\ref{finalFig}).

In this work, we investigated a general class of non-linear voter models of
opinion dynamics in one dimension.  We considered the situation where the
transition rate for each voter depends in a non-trivial way on the number of
disagreeing neighbors.  In general, the average magnetization is not
conserved in these models and the evolution of the average opinion is coupled
to higher-order opinion correlations.  It is possible to truncate the
hierarchy of equations for these correlations in a simple and plausible
manner.  From this approach, we find that the system coarsens, albeit
differently than in the pure voter model because of the interactions
(repulsion or attraction) of neighboring domain walls.  The probability to
reach the final state of $\uparrow$ consensus is also shown to have a
non-trivial initial state dependence, a feature that reveals the tendency of
the system to inhibit or to reach consensus.  The decoupling approximation
presented in this letter appears to be both efficient and robust, and
therefore it could well be useful in other models of opinion formation,
language dynamics, {\it etc.} \cite{review}, where the interactions between
agents are not conservative.

\acknowledgments We gratefully acknowledge the support of the ARC ``Large
Graphs and Networks'' (RL), NSF grant DMR0535503 (SR), and the hospitality of
the Ettore Majorana Center where this project was initiated.

{\it Added Note:} After this manuscript was completed, we became aware of
parallel work by Slanina et al.\ \cite{slanina}, in which they obtain results
similar to ours.

\end{document}